\documentclass{iopart}
\usepackage{iopams}
\usepackage{graphicx,epsf}
\usepackage{color}
\usepackage{mathrsfs}
\usepackage{dcolumn}
\usepackage{cite}

\usepackage{subfig}

\begin{document}


\title[]{Evidence of ``Two Plasmon" Decay of Energetic Particle Induced Geodesic Acoustic Mode}

\author{Zhiyong Qiu$^{1,2}$, Liu Chen$^{1,2,3}$, Fulvio Zonca$^{2,1}$ and Matteo Valerio Falessi$^{2}$}

\address{$^1$Institute for    Fusion Theory and Simulation and Department of Physics, Zhejiang University, Hangzhou, P.R.C}
\address{$^2$ Center for Nonlinear Plasma Science and ENEA C. R. Frascati, Frascati, Italy}
\address{$^3$Department of   Physics and Astronomy,  University of California, Irvine CA 92697-4575, U.S.A.}

\begin{abstract}
Secondary low frequency mode generation by    energetic particle induced geodesic acoustic mode (EGAM) observed in LHD experiment is studied using nonlinear gyrokinetic theory. It is found that  the   EGAM frequency can be significantly higher than local geodesic acoustic mode (GAM) frequency in low collisionality plasmas, and  it can decay into two GAMs as its frequency approaches twice GAM frequency, in a process analogous to the well-known two plasmon decay instability. The condition for this process to occur is also discussed.
\end{abstract}

\maketitle

\section{Introduction}

Confinement of plasmas in magnetically confined fusion devices \cite{KTomabechiNF1991}  is one of  the key issues for the sustained burning and fusion gain. The anomalous transport induced by micro-scale turbulence   excited by expansion free energy intrinsic to confinement is thus an important topic of fusion plasma research \cite{WhortonRMP1999,LChenJGR1999}. Zonal field structures (ZFS) with toroidally and nearly poloidally symmetric mode structures ($n=0/m\simeq0,\pm1\cdots$, with  $n/m$ being the toroidal/poloidal mode numbers, respectively)   are generally recognized to regulate micro-scale drift wave  turbulence (DW) including drift Alfv\'en waves (DAWs) and their associated transport by scattering into stable short radial wavelength regimes \cite{AHasegawaPoF1979,ZLinScience1998,ADimitsPoP2000,LChenPoP2000,PDiamondPPCF2005}.  Interested readers may refer to a recent work  \cite{FZoncaJPCS2021} discussing the role of ZFS  and  phase space zonal structures (PSZS) \cite{FZoncaNJP2015,MFalessiPoP2019} as the generator of nonlinear equilibria with
(suppressed) turbulence \cite{ADimitsPoP2000,LChenNF2007a}.

Geodesic acoustic modes (GAMs) \cite{NWinsorPoF1968,FZoncaEPL2008} are ZFS unique in toroidal plasmas;  oscillating at a finite frequency in the ion sound frequency regime due to toroidicity induced plasma compression. GAMs are predominantly electrostatic  radial corrugations with the scalar potential characterized by  $n=0/m\simeq0,\pm1,\cdots$, and an    up-down anti-symmetric $m\simeq1$ density perturbation.  Consequently, GAM, being toroidally symmetric,  cannot  by itself induce  cross-field transport. Instead,   GAMs can  regulate  DW/DAWs and the associated transport  via  spontaneous excitation, since they can scatter DW/DAWs into stable short radial wavelength domain \cite{NChakrabartiPoP2007,FZoncaEPL2008,ZQiuPoP2014,ZQiuNF2019b}. Interested readers may refer to Ref. \cite{ZQiuPST2018} for a brief review of kinetic theories of GAM, including linear dispersion relation, excitation by super-thermal  energetic particles (EPs) and nonlinear interaction with DWs/DAWs.

Due to their finite frequencies, GAMs can resonantly interact and exchange energy with charged particles including super-thermal EPs,  leading to, respectively, collisionless Landau damping \cite{HSugamaJPP2006,ZQiuPPCF2009} and  resonant excitation by EPs \cite{HBerkNF2006,RNazikianPRL2008,TIdoNF2015}.  The excitation of EP-induced GAM (EGAM) was first analytically investigated in Ref. \cite{GFuPRL2008}, showing the dominant role played by wave-particle resonant interactions and free energy associated with EP velocity space anisotropy,      with application to different scenarios characterized by different EP distribution functions \cite{HBerkNF2010,ZQiuPPCF2010,HWangPRL2013,JBaptistePoP2014,HWangPoP2015,JCaoPoP2015}.  The global features due to GAM continuum  coupling are also investigated \cite{FZoncaIAEA2008},  yielding a finite threshold due to continuum damping, or mode conversion to propagating kinetic GAM, as kinetic dispersiveness of both thermal and energetic ions are properly accounted for \cite{ZQiuPPCF2010,ZQiuPoP2012}.  The EGAMs are typically characterized by a global mode structure with frequency lower than local GAM frequency and  radial extension determined by EP density profile and kinetic dispersiveness, $L\sim \sqrt{L_n\rho_h}$, with $L_n$ being the EP density profile scale length, and $\rho_h$ being EP characteristic orbit width. Nonlinear saturation of EGAM due to wave-particle trapping \cite{TOneilPoF1965} in the weak  drive limit is investigated \cite{ZQiuPST2011,ABiancalaniJPP2017}, and the EGAM induced EP loss via pitch angle scattering into unconfined orbits is presented in Ref. \cite{DZarzosoNF2018}.

Recently, a peculiar phenomena was reported in Large Helical Device (LHD) low density  experiments with neutral beam injection (NBI) heating \cite{TIdoNF2015}. During the discharge with relatively high plasma temperature (central electron temperature up to $\sim$ 7keV) and   low plasma density (line averaged density $\sim 0.1\times 10^{19} m^{-3}$), the EGAM frequency is observed to be significantly higher than local GAM frequency. The interpretation was given in Ref. \cite{JCaoPoP2015}, where it was shown that, due to the low collisionality in the high-temperature/low-density discharges, the injected beam ions are not fully slowed down, and form a bump-on-tail type EP distribution function. As a consequence,  the free energy associated with the positive slope in the low energy side of the distribution function, provides an additional drive, that generates a new unstable branch with the frequency being significantly higher than local GAM frequency \cite{JCaoPoP2015}.  A similar interpretation was also provided in Ref. \cite{HWangPoP2015} considering a similar distribution function induced by low charge exchange rate.

It was further found \cite{TIdoPRL2016} that, as the high frequency EGAM (will be denoted as ``primary EGAM" in the rest of the paper for apparent reasons) chirped up to twice local GAM frequency, possibly as a result of the EGAM induced pitch angle scattering to lower pitch angle and thus higher $v_{\parallel}$ domain,  a ``secondary mode" could be strongly driven unstable, with the frequency close to the local GAM frequency \cite{TIdoPRL2016}. The experimental observations are nicely recovered in a MHD-kinetic hybrid simulation using MEGA code \cite{YTodoNF2010,HWangPRL2018}, and it is found that, besides the secondary mode strong  excitation,  the EPs ``driving" the secondary mode are the same as those linearly driving the primary EGAM, though the secondary mode frequency is only half of that of the primary mode. It is also found that  the secondary mode generation  still persists when the ``fluid nonlinearity" is turned off; and, as the primary mode frequency keeps chirping up, the secondary mode frequency is almost unchanged, suggesting it is a normal mode of the system itself. One speculation by Ref. \cite{HWangPRL2018} is that the secondary mode is driven by ``nonlinear resonance" \cite{GKramerPRL2012,LChenPST2019}, which, however,  is typically associated with finite amplitude fluctuations, and is not satisfied in  the condition for the onset of the  secondary mode.

In this work, we will show  that, the mechanism underlying the  secondary mode generation observed in Ref. \cite{TIdoPRL2016} is analogous to the well-known two plasmon decay process \cite{CSLiuPoF1976,LPowersPoF1984}, where an incident electromagnetic wave decays into two plasma waves identical to each other.
The primary mode corresponds to the linearly unstable high frequency EGAM given in Ref. \cite{JCaoPoP2015}, while  the secondary mode corresponds to a linearly stable branch described  by the same linear dispersion relation, with the frequency determined by the local GAM frequency. The secondary mode can be nonlinearly driven unstable as the primary mode frequency approaches twice the secondary mode frequency (which is close to the local GAM frequency), which   minimizes the threshold due to frequency mismatch.
The interpretation is consistent with  the crucial elements from experimental observation \cite{TIdoPRL2016} and numerical simulation  \cite{HWangPRL2018};  thus, it provides and illuminates the underlying physics picture.

The rest of the paper is organized as follows. In Sec. \ref{sec:model}, the theoretical model is given. The linear particle response  to  EGAM is derived in Sec. \ref{sec:linear_property}, where the linear EGAM properties in the LHD low collisionality plasma are  also reviewed. In Sec. \ref{sec:two_plasmon_decay}, the nonlinear decay of the primary mode into two low frequency GAMs is investigated, and its correspondence to the experimental observations and MEGA simulations are discussed. Finally, a brief summary and discussion is presented in Sec. \ref{sec:summary}. For the self-containedness of the materials, the properties of the three branches of the linear dispersion relation are briefly reproduced in   \ref{sec:app_A}, while the frequency up-chirping of the primary EGAM, being an important condition for the decay process, is briefly outlined in \ref{sec:app_B}.

\section{Theoretical Model}
\label{sec:model}

In the above described nonlinear decay process, a primary EGAM ($\Omega_0\equiv\Omega_0(\omega_0,k_{r,0})$) decays into two secondary modes with almost identical frequency, $\Omega_1\equiv\Omega_1(\omega_1,k_{r,1})$ and $\Omega_2\equiv\Omega_2(\omega_2,k_{r,2})$. All the three modes, are  EGAM/GAMs with $n=0/m\simeq0$, thus the toroidal/poloidal wavenumber matching condition is naturally satisfied, and only the constraint on frequency and radial wavenumber matching condition is needed. For the modes with
predominantly electrostatic polarization,  the
equations describing the nonlinear decay of the linearly unstable primary EGAM, can be derived from the charge quasi-neutrality condition. Assuming for simplicity the bulk ions and EPs are the same ion species with unit electric charge, the quasi-neutrality condition can be written as:
\begin{eqnarray}
\delta n_e=\delta n_i+\delta n_h,\label{eq:QN}
\end{eqnarray}
with the perturbed particle density derived from the perturbed distribution function as
\begin{eqnarray}
\delta n_s\equiv \left\langle (q/m)\partial_E F_{0,s}\delta\phi+J_0(k_{\perp}\rho_{L,s})\delta H_s\right\rangle.
\end{eqnarray}
Here, subscripts $s=e,i,h$ denote electrons, ions and energetic (hot) particles, respectively, $F_{0,s}$ is the equilibrium distribution function, $\partial_E F_{0,s}$ is the short notation for $\partial F_{0,s}/\partial E$, with $E\equiv v^2/2$,  $J_0(k_{\perp}\rho_{L,s})$ is the Bessel function of zero index accounting for finite Larmor radius (FLR) effects, $k_{\perp}$ is the perpendicular wavenumber,   $\rho_L\equiv v_{\perp}/\Omega_s$ is the   Larmor radius with $\Omega_s$ being gyro-frequency, and $\langle\cdots\rangle$ denotes velocity space integration.  $\delta H$ is the non-adiabatic particle response to GAM/EGAM, and can be derived from nonlinear gyrokinetic equation \cite{EFriemanPoF1982}:
\begin{eqnarray}
\left(\partial_t+\omega_{tr}\partial_{\theta}+i\omega_d \right)\delta H_k&=&-i\omega\frac{q}{m}\partial_E F_0J_k\delta\phi_k\nonumber\\ &&-\frac{c}{B_0}\sum_{\mathbf{k}=\mathbf{k'}+\mathbf{k''}} \hat{\mathbf{b}}\cdot\mathbf{k}''\times\mathbf{k}' J_{k'}\delta\phi_{k'}\delta H_{k''}.\label{eq:GKE}
\end{eqnarray}
Here, $\omega_{tr}\equiv v_{\parallel}/(qR_0)$ is the transit frequency, $\omega_d=k_r v_d$ is the magnetic drift frequency associated with geodesic curvature, and $v_d=(v^2_{\perp}+2 v^2_{\parallel})\sin\theta/(2 \Omega R_0)\equiv \hat{v}_d\sin\theta$ is the magnetic   drift velocity, $J_k\equiv J_0(k_\perp\rho_L)$ for simplicity of notation, and $\sum_{\mathbf{k}=\mathbf{k'}+\mathbf{k''}}$ denotes the usual selection rule for frequency/wavenumber matching condition for the nonlinear mode coupling. The other notations are standard.

The   thermal plasma linear response to GAM/EGAM, can be readily obtained from Ref. \cite{ZQiuPPCF2009}, and  after surface averaging, equation (\ref{eq:QN}) reduces to
\begin{eqnarray}
-\frac{e}{m_i}n_0k^2_r\frac{1}{\Omega^2_i}\left(1-\frac{\omega^2_G}{\omega^2}\right)\overline{\delta\phi}_k+\overline{\delta n}^{NL}_i + \overline{\delta n}_h=0, \label{eq:QN_EGAM}
\end{eqnarray}
with the leading order linear thermal plasma response properly accounted for by the first term, from which the linear dispersion relation of GAM can be obtained, and $\overline{(\cdots)}\equiv\int^{2\pi}_0(\cdots)d\theta/(2\pi)$ is the surface averaging. Furthermore, $\omega_G\equiv \sqrt{7/4+\tau} v_{it}/R$ is the leading order  GAM frequency with $\tau=T_e/T_i$ being the temperature ratio and $v_{it}$ being ion thermal velocity, while higher order terms such as FLR and/or finite parallel compression can be straightforwardly accounted for by replacing $\omega_G$ with more accurate expression \cite{ZQiuPPCF2009}.    The linear EGAM dispersion relation can be derived by keeping $\delta n^L_h$ in equation (\ref{eq:QN_EGAM}), with the free energy from EP velocity space anisotropy  and the characteristic features of the EGAM determined by the specific EP distribution function \cite{GFuPRL2008,HBerkNF2010,ZQiuPPCF2010,JCaoPoP2015,HWangPoP2015}. Nonlinear modulation of GAM/EGAM by  thermal plasma nonlinearity, can be accounted for by $\delta n^{NL}_i$, including excitation by DW/DAWs \cite{FZoncaEPL2008,ZQiuEPL2013,ZQiuPoP2014}.   Here, super-scripts ``L" and ``NL" represent linear and nonlinear responses, respectively.

For the LHD low collisionality discharge \cite{TIdoNF2015} of interest, where plasma density is very low while the electron temperature is very high, the EP is characterized by a not fully slowed down distribution function, analogous to a bump-on-tail case, and the corresponding linear properties of the EGAM are investigated in Ref. \cite{JCaoPoP2015}. The linear EGAM properties are the basis of the present nonlinear analysis, and, thus, for the self-containedness  of the present nonlinear analysis, the results of Ref. \cite{JCaoPoP2015} will be briefly summarized in Sec. \ref{sec:linear_property}

In this work, both the thermal plasma and EP induced nonlinear coupling are consistently derived \cite{LChenEPL2014,ZQiuPoP2017}, by including their nonlinear contribution to density perturbation in the quasi-neutrality condition. As we will show \textit{a posteriori}, the nonlinear coupling is dominated by the EP finite orbit width effects (FoWs), with resonant EPs playing the crucial role \cite{GFuJPP2011,ZQiuPoP2017}.  The thermal plasma contribution to the nonlinear coupling \cite{LChenEPL2014},  meanwhile,  will be shown to be negligible compared    to the dominant role of EPs.

\section{Linear properties}
\label{sec:linear_property}

For the completeness of this work, we will briefly derive the linear EP response to GAM/EGAM, which will be used in deriving the nonlinear response of EP to the secondary EGAMs.  Separating the linear from nonlinear responses by taking $\delta H_k=\delta H^L_k+\delta H^{NL}_k$,    the linear EP response to EGAM can be derived from the linear gyrokinetic equation,
\begin{eqnarray}
\left(\partial_t+\omega_{tr}\partial_{\theta}+i\omega_d\right)\delta H^L_k=-i\omega\frac{e}{m}\partial_E F_0J_k\delta\phi_k. \label{eq:GKE_L}
\end{eqnarray}
Equation (\ref{eq:GKE_L}) can be solved by transforming into the EP drift orbit center coordinate by taking $\delta H^L_k\equiv e^{i\Lambda_k} \delta H^L_{dk}$, with $\Lambda_k$ satisfying  $\omega_{tr}\partial_{\theta}\Lambda_k+\omega_{d,k}=0$. Here, for simplicity of discussion and focus on proof of principle demonstration, well circulating EPs are assumed, thus, variation of $v_{\parallel}$  along the magnetic field is neglected,  and one has $\Lambda_k=\hat{\Lambda}_k\cos\theta$, with $\hat{\Lambda}_k=k_r\hat{\rho}_d$ being radial wave-number  normalized to drift orbit width, and $\hat{\rho}_d = \hat{v}_{dr}/\omega_{tr}$ is the EP magnetic drift orbit width. The generalization to finite inverse aspect ratio case as well as general geometry and particle orbits is straightforward.  Furthermore, $\tau\equiv T_e/T_i\ll1$ is assumed for simplicity, such that one has $\delta\phi_G\simeq \overline{\delta\phi}_G$ while $\omega_{tr,e}\gg\omega_G$ is still satisfied \footnote{Interested readers may refer to Ref. \cite{ZQiuPPCF2009} for the contribution of finite $\tau$ to linear GAM dispersion relation  via the $m\neq0$ components of the scalar potential.}.    We then have,
\begin{eqnarray}
\left(\partial_t+\omega_{tr}\partial_{\theta}\right)\delta H^L_{dk}=- \frac{e}{m} e^{-i\hat{\Lambda}_k\cos\theta}J_k\partial_t \overline{\delta\phi}_k\partial_E F_{0,h},
\end{eqnarray}
from which   $\delta H^L_{dk}$ can be derived as
\begin{eqnarray}
\delta H^L_{dk}=-\frac{e}{m}\partial_E F_{0,h}J_k\overline{\delta\phi}_k\sum^{\infty}_{l=-\infty} \frac{\omega}{\omega-l\omega_{tr}} (-i)^l J_l(\hat{\Lambda}_k) e^{il\theta}.\label{eq:EP_drift_center}
\end{eqnarray}
Here, $l$ is integer, and $\sum_l\equiv \sum_{l=-\infty}^{\infty}$ will be used later for simplicity of notation. In deriving equation (\ref{eq:EP_drift_center}), the Jacobi-Anger expansion $e^{i\hat{\Lambda}_k\cos\theta}=\sum_l i^l J_l(\hat{\Lambda}_k)e^{il\theta}$ is used.  Pulling back to the EP guiding center coordinate, we then have, the linear well-circulating EP response to EGAM
\begin{eqnarray}
\delta H^L_{k}=-\frac{e}{m}\partial_E F_{0,h}J_k\overline{\delta\phi}_k\sum_{l,p} \frac{\omega}{\omega-l\omega_{tr}} i^{-l+p}J_l(\hat{\Lambda}_k) J_p(\hat{\Lambda}_k) e^{i(l+p)\theta}.\label{eq:EP_linear}
\end{eqnarray}

Substituting equation (\ref{eq:EP_linear}) into the surface averaged quasi-neutrality condition, equation (\ref{eq:QN_EGAM}), one then obtains,
\begin{eqnarray}
 \frac{e}{m_i}n_0k^2_r\frac{1}{\Omega^2_i}\mathscr{E}_{EGAM} \overline{\delta\phi} =0, \label{eq:DR_linear_EGAM}
\end{eqnarray}
with the EGAM dispersion relation given by
\begin{eqnarray}
\mathscr{E}_{EGAM}\equiv&& -1+\frac{\omega^2_G}{\omega^2}\nonumber\\
&&+ \frac{\sqrt{2}\pi B_0e^2}{n_0} \int\frac{(2-\lambda B_0)^2}{(1-\lambda
B_0)^{1/2}}\frac{E^{5/2} dEd\lambda \partial_EF_{0,h}}{2E(1-\lambda
B_0)-\omega^2q^2R^2_0}. \label{eq:EGAM_DR}
\end{eqnarray}
Here, only the $l=\pm1$ transit harmonic are kept in equation (\ref{eq:EGAM_DR}), in consistency with the typical $\hat{\Lambda}_k\ll1$ ordering for EGAMs with global mode structure.
The EGAM stability  depends sensitively on the  specific EP equilibrium distribution function.   For the  not fully slowed down EPs in LHD experiments \cite{TIdoNF2015} due to low collisionality, the distribution can be modelled as
\begin{eqnarray}
F_{0,h}=\frac{c_0H(E_b-E)H(E-E_L)}{E^{3/2}+E^{3/2}_{crit}}\delta(\lambda-\lambda_0),\label{eq:EP_DF}
\end{eqnarray}
which can be solved as a dynamic  solution of the Fokker-Planck equation with the collisional operator dominated by thermal electron induced slowing down \cite{JCaoPoP2015}. Here, $\lambda\equiv \mu B_0/E$ is the pitch angle, $\mu\equiv m v^2_{\perp}/B$ is the magnetic moment,   $H(\lambda-\lambda_0)$ is the Heaviside step function,  $E_b$ is the EP birth energy, $E_L\simeq E_b\exp(-2\gamma_c t)$ is the time dependent  lower end of the distribution function,   and $E_{crit}$ is the critical energy at which the EP pitch angle scattering rate off thermal ions is comparable with the   slowing down rate $\gamma_c$ \cite{TStixPoP1972}. The normalization to EP density $n_b$ gives  $c_0=n_b \sqrt{1-\lambda_0B_0}/(2\sqrt{2}\pi B_0\ln (E_b/E_L))$.  Note that, $c_0$ is proportional to $\Gamma/\gamma_c$
 with $\Gamma$ being the NBI particle flux and  $\gamma_c$ is the slowing-down rate on thermal electrons.    The resulting dispersion relation is given as \cite{JCaoPoP2015}
\begin{eqnarray}
&&-1+\frac{\omega^2_G}{\omega^2}+N_b \left[C\left(\ln\left(1-\frac{\omega^2_b}{\omega^2}\right) - \ln\left(1-\frac{\omega^2_L}{\omega^2}\right)\right)\right.\nonumber\\
&&\left.+D\left(\frac{1}{1-\omega^2_b/\omega^2}-\frac{1}{1-\omega^2_L/\omega^2}\right)\right]=0.\label{eq:EGAM_DR_linear}
\end{eqnarray}
Here, $N_b\equiv n_b q^2\sqrt{1-\lambda_0B_0}/(4\ln (E_b/E_L) n_0)$ is the  ratio of EP to bulk plasma density, $C=(2-\lambda_0B_0)(5\lambda_0B_0-2)/(2(1-\lambda_0B_0)^{5/2})$, $D=\lambda_0B_0(2-\lambda_0B_0)^2/(1-\lambda_0B_0)^{5/2}$, and $\omega_b\equiv \sqrt{2E_b(1-\lambda_0B_0)}/(qR_0)$ and $\omega_L\equiv \sqrt{2E_L(1-\lambda_0B_0)}/(qR_0)$ are the transit frequencies defined at  $E_b$ and   $E_L$, respectively.  The first term in the square bracket corresponds to the slowing-down distribution induced logarithmic singularity, and it is destabilizing for $\lambda_0B_0>2/5$ \cite{ZQiuPPCF2010}; while the single pole like singularity in the second term is from the low energy side cutoff, and it is always destabilizing \cite{JCaoPoP2015}.

The dispersion relation (\ref{eq:EGAM_DR_linear})  have  three branches, i.e.,  an unstable branch determined by and close to $\omega_L$,   and is denoted as ``lower beam branch" (LBB);  a   stable branch determined by the local GAM frequency,  denoted as the ``GAM branch" (GB), and is little affected by the EP distribution function; and a marginally stable branch determined by the birth energy of the distribution function ($E_b$), and is denoted as ``upper beam branch" (UBB). It is noteworthy that, the unstable LBB   can have a frequency much higher than the local GAM frequency, while $\gamma/\omega_G\propto   \sqrt{N_b}$ can be obtained by balancing the thermal plasma contribution to the dominant destabilizing term due to low energy end cutoff (i.e., the term proportional to $1/(1-\omega^2_L/\omega^2)$ in equation (\ref{eq:EGAM_DR_linear})).  The stability of the three branches described by the dispersion relation (\ref{eq:EGAM_DR_linear}) on $\omega_L$   is     briefly summarized in \ref{sec:app_A} for the convenience of readers. One important feature is that, the linear unstable LBB frequency is non-perturbatively determined by $\omega_L$, and thus, will self-consistently chirp up or down, due to EGAM induced pitch angle scattering, as we show in   \ref{sec:app_B}.

Thus, the primary mode in LHD experiment \cite{TIdoNF2015} corresponds to the linearly unstable LBB, while the secondary mode with the frequency being local GAM frequency, corresponds to the linearly stable GB.
We will show in the next section that, as the   unstable LBB  chirping up to twice local GAM frequency, it can decay into two linearly stable GBs, similar to the well-known ``two plasmon decay" process describing an electromagnetic  wave decay into two Langmuir waves \cite{CSLiuPoF1976,LPowersPoF1984}.

\section{Two plasmon decay of EGAM}
\label{sec:two_plasmon_decay}

In this section, the high frequency LBB decay into two linearly stable GBs, will be investigated using nonlinear  gyrokinetic theory.  Denoting the ``primary mode"  and  two ``secondary modes"  with subscripts ``0", ``1" and ``2", respectively, the nonlinear particle responses to GAM/EGAM can be derived from nonlinear gyrokinetic equation
\begin{eqnarray}
\left(\partial_t +\omega_{tr}\partial_{\theta}+i\omega_d\right)\delta H^{NL}_k=-\frac{c}{B_0}\sum_{\mathbf{k}=\mathbf{k}''+\mathbf{k}'} \hat{\mathbf{b}}\cdot\mathbf{k}''\times\mathbf{k'} J_{k'}\overline{\delta\phi}_{k'}\delta H_{k''}.
\end{eqnarray}
To properly assess the  nonlinear coupling, the contribution of both thermal plasmas and EPs are considered simultaneously, by deriving their nonlinear response to GAM/EGAM, and evaluating their contribution to the surface averaged quasi-neutrality condition, equation (\ref{eq:QN_EGAM}).

\subsection{Negligible thermal plasma contribution to nonlinear coupling}

For electrons with $\omega_{tr,e}\gg\omega_G$, the nonlinear electron response is dominated by surface averaged contribution. Noting $\delta H^L_e = (e/T_e)F_{0,e}\overline{\delta\phi}_k$,  one has
\begin{eqnarray}
\partial_t\delta H^{NL}_{k,e} = -\frac{c}{B_0}\sum_k \hat{\mathbf{b}}\cdot\mathbf{k''}\times\mathbf{k'} \overline{\delta\phi}_{k'}\frac{e}{T_e}F_{0,e}\overline{\delta\phi}_{k''}=0,
\end{eqnarray}
i.e., $\delta H^{NL}_{k,e}=0$ up to the $O(\omega_G/\omega_{tr,e})\ll1$  order. Thus, electron contribution to the nonlinear coupling can be neglected.

Nonlinear ion response can be derived, noting the $\omega_G\gg\omega_{tr,i}, \omega_d$ ordering  and  that $\delta H^L_i\simeq -(e/T_i) J_0\overline{\delta\phi}_G (1+\omega_d/\omega)$, and one has
\begin{eqnarray}
\partial_t\delta H^{NL}_{k,i}&\simeq& -\frac{c}{B_0}\sum_k  \hat{\mathbf{b}}\cdot\mathbf{k''}\times\mathbf{k'} J_{k'} \overline{\delta\phi}_{k'} \delta H_{k''}\nonumber\\
&\simeq& -\frac{c}{B_0} \left[ J_{k'}\partial_r \overline{\delta\phi}_{k'}\frac{1}{r}\partial_{\theta}\delta H^L_{i,k''}- J_{k''}\partial_r \overline{\delta\phi}_{k''}\frac{1}{r}\partial_{\theta}\delta H^L_{i,k'} \right]. \label{eq:ion_contribution_couple}
\end{eqnarray}
Note that, for GAM/EAM with $n=0$,   the geodesic curvature induced drift $\omega_d\propto\sin\theta$, which gives $\delta H^{NL}_{k,i}$ a $\propto \cos\theta$ dependence to the leading order. Thus, finite contribution to   the surface averaged quasi-neutrality can only enter  through toroidal coupling ($B\propto 1-\epsilon\cos\theta$), as was discussed in Ref. \cite{HZhangNF2009}.   Thus, the thermal ion induced nonlinearity via surfaced averaged quasi-neutrality condition, can be estimated to be of   order  $\sim c\epsilon n_0 k_r \delta\phi_0\delta\phi_1  \omega_{d,i}/(\omega_G r B_0)$,    which is comparable to parallel nonlinearity  and is, thus, negligibly small. Consequently, we expect the EP induced nonlinear coupling will be dominant, as shown in Refs. \cite{GFuJPP2011,ZQiuPoP2017}.

\subsection{Finite EP contribution to nonlinear coupling via EP FoW effects}

Nonlinear EP response to the sideband $\Omega_k$   can be derived from nonlinear gyrokinetic equation, by drift orbit center coordinate transformation. For the $\Omega_1$ generation due to the coupling of $\Omega_0$ and $\Omega_{2^*}$ coupling,
taking $\delta H^{NL}_1=e^{i\Lambda_1}\delta H^{NL}_{d1}$, the corresponding   equation for nonlinear EP response to $\Omega_1$  can be written as
\begin{eqnarray}
\hspace*{-2em}\left(\partial_t+\omega_{tr}\partial_{\theta}\right)\delta H^{NL}_{d1}&&=-\frac{c}{B_0} e^{-i \Lambda_1}\sum_{\mathbf{k}_1}  \hat{\mathbf{b}}\cdot\mathbf{k''}\times\mathbf{k'} J_{k'}\overline{\delta\phi}_{k'}\delta H^L_{h,k''}\nonumber\\
&&=-\frac{c}{r B_0}e^{-i \Lambda_1}\left[\partial_r\overline{\delta \phi}_{2^*} \partial_{\theta}\delta H^L_0- \partial_r\overline{\delta \phi}_0 \partial_{\theta}\delta H^L_{2^*} \right].\label{eq:GKE_NL_DC}
\end{eqnarray}
Here, one expects that the contribution from  $\overline{\delta\phi}_{2^*}\delta H^L_0$ being dominant with respect to $\overline{\delta\phi}_0\delta H^L_{2^*}$  due to the crucial contribution of resonant EPs (note that $\Omega_0$ is linearly unstable due to resonant EP drive). However, both terms are consistently kept for now, with both resonant and non-resonant EP contribution to the nonlinear coupling accounted for on the same footing.
As we will show \textit{a posteriori}, the nonlinear coupling is dominated by EP FoW effects, while the  sub-dominant EP FLR effects can be neglected systematically.  Substituting the linear EP response into equation (\ref{eq:GKE_NL_DC}), we then have,
\begin{eqnarray}
\delta H^{NL}_{d,1}&=&-\frac{c}{rB_0}\frac{e}{m}\partial_E F_{0,h}\sum_{l,p,\eta} J_{\eta}(\hat{\Lambda}_1) i^{-\eta}e^{i\eta\theta} \nonumber\\
&\times&\left[\partial_r\overline{\delta\phi}_{2^*}\overline{\delta\phi}_0 \frac{i(l+p)\omega_0}{\omega_0-l\omega_{tr}}  \frac{J_l(\hat{\Lambda}_0)J_p(\hat{\Lambda}_0)}{\omega_1-(l+p+\eta)\omega_{tr}} i^{-l+p} e^{i(l+p)\theta}\right.\nonumber\\
&-&\left. \partial_r\overline{\delta\phi}_0\overline{\delta\phi}_{2^*}\frac{i(l+p)\omega_{2^*}}{\omega_{2^*}-l\omega_{tr}} \frac{J_l(\hat{\Lambda}_2)J_p(\hat{\Lambda}_2)}{\omega_1-(-l+p+\eta)\omega_{tr}} i^{l-p}  e^{i(-l+p)\theta} \right].\nonumber
\end{eqnarray}

 The nonlinear EP response to $\Omega_1$ can then be obtained, by the pull-back transformation, and one has
\begin{eqnarray}
\hspace*{-2em}\delta H^{NL}_{1}&=&-i\frac{c}{rB_0}\frac{e}{m}\partial_E F_{0,h}\sum_{l,p,\eta,h} J_{\eta}(\hat{\Lambda}_1) J_{h}(\hat{\Lambda}_1) i^{-\eta+h}e^{i(\eta+h)\theta} \nonumber\\
&\times&\left[\partial_r\overline{\delta\phi}_{2^*}\overline{\delta\phi}_0 \frac{(l+p)\omega_0}{\omega_0-l\omega_{tr}}  \frac{J_l(\hat{\Lambda}_0)J_p(\hat{\Lambda}_0)}{\omega_1-(l+p+\eta)\omega_{tr}} i^{-l+p} e^{i(l+p)\theta}\right.\nonumber\\
&-&\left. \partial_r\overline{\delta\phi}_0\overline{\delta\phi}_{2^*}\frac{(l+p)\omega_{2^*}}{\omega_{2^*}-l\omega_{tr}} \frac{J_l(\hat{\Lambda}_2)J_p(\hat{\Lambda}_2)}{\omega_1-(-l+p+\eta)\omega_{tr}} i^{l-p}  e^{i(-l+p)\theta} \right].\label{eq:EP_NL_EGAM}
\end{eqnarray}
In the above expression, the first term in the bracket comes from $\overline{\delta\phi}_{2^*}\delta H^L_0$, as evident from the denominator $\omega_0-l\omega_{tr}$, while the other term comes from $\overline{\delta\phi}_0\delta H^L_{2^*}$.  For simplicity, in the following derivation, only the first term due to $\overline{\delta\phi}_{2^*}\delta H^L_0$ will be kept, which can be dominant due to resonant EP contribution. The other term, can also contribute and quantitatively impact the nonlinear process, but we expect its contribution is relatively small.
The physics meaning of various terms in equation (\ref{eq:EP_NL_EGAM}) is clear, in that  $\omega_0-l\omega_{tr}$  in  the denominator  gives wave-particle power exchanges with the pump $\Omega_0$, $(l+p)$ comes from $\partial_{\theta}\delta H^L_0$ in the perpendicular nonlinearity, and the Bessel functions are from EP FoW effects, determining the strength  of EPs interaction  with the mode.

For finite $\Omega_1$ generation due to $\overline{\langle\delta H^{NL}_1 \rangle}$,  assuming $|\hat{\Lambda}_k|\ll1$ for EGAMs with typically global mode structure, one requires the following ``selection rules" for non-vanishing EP contribution: 1. $l+p+\eta+h=0$ for non-vanishing component   surface average, 2. $l+p\neq 0$ for $\partial_{\theta}\delta H^L_0\neq0$, 3. $l\neq0$ for finite EP linear drive to the primary EGAM, and 4. $|l|+|p|+|\eta|+|h|$ as small as possible for maximized contribution under the $\hat{\Lambda}_k\ll1$ assumption.  One then has, the dominant contribution comes from $l=1, p=0, \eta=-1, h=0$ or $l=1,p=0,\eta=0, h=-1$, which gives:
\begin{eqnarray}
\overline{\delta H^{NL}_1}=&-&\frac{c}{rB_0} \frac{e}{m} \partial_E F_{0,h}  \partial_r\overline{\delta\phi}_{2^*} \overline{\delta\phi}_0 \frac{\omega_0}{\omega_0-\omega_{tr}}\frac{\omega_{tr}}{\omega_1(\omega_1-\omega_{tr})} \nonumber\\ &\times&J_1(\hat{\Lambda}_0)J_0(\hat{\Lambda}_0)J_1(\hat{\Lambda}_1) J_0(\hat{\Lambda}_1). \label{eq:EP_couple}
\end{eqnarray}

The  ratio of  thermal ion and EP contribution to the  nonlinear coupling,  obtained from equations (\ref{eq:ion_contribution_couple}) and (\ref{eq:EP_couple}), respectively,  can be estimated to be
$\mathscr{N}_i/\mathscr{N}_h\sim  [n_0 \epsilon \omega_{d,i}/\omega_G]/[N_b (\omega_0/\gamma)(\omega_{d,h}/\omega_{tr,h})^2] \sim \sqrt{N_b}\epsilon/\sqrt{\Lambda_h}\ll1$; confirming the conjecture following equation (\ref{eq:ion_contribution_couple}). Here, $\mathscr{N}_i$ and $\mathscr{N}_h$ are, respectively, the thermal ion and EP contribution to nonlinear coupling, respectively. Linear EGAM orderings, including  $\beta_h\lesssim\beta_i$, $|\gamma/\omega_G|\sim \sqrt{N_b}$ and $\omega_{tr,h}\sim \omega_G$ \cite{ZQiuPPCF2010} are used in estimating the ordering of $\mathscr{N}_i/\mathscr{N}_h$.

Substituting equation (\ref{eq:EP_couple}) into quasi-neutrality condition, one obtains  the equation for $\Omega_1$ generation due to $\Omega_0$ and $\Omega_{2^*}$ coupling
\begin{eqnarray}
\mathscr{E}_1 \overline{\delta\phi}_1&=& \mathscr{A} \mathscr{D}_1  \frac{\omega_0}{\omega_1k^2_{r,1}} \partial_r\overline{\delta\phi}_{2^*}\overline{\delta\phi}_0. \label{eq:DR_1}
\end{eqnarray}
Here, $\mathscr{E}_1\equiv\mathscr{E}_{EGAM} (\omega_1, k_{r,1})$ is the linear EGAM dispersion relation derived in Ref. \cite{JCaoPoP2015}, which is linearly stable for the GB $\Omega_1$,   with the frequency determined by GAM frequency. Furthermore, $\mathscr{A}\equiv c\Omega^2_i/(rB_0n_0)$, and
\begin{eqnarray}
\mathscr{D}_1\equiv \left\langle \frac{ \partial_E F_{0,h}}{\omega_0-\omega_{tr}}  \frac{\omega_{tr}}{\omega_1-\omega_{tr}} J_0(\hat{\Lambda}_0)J_1(\hat{\Lambda}_0)J_0(\hat{\Lambda}_1) J_1(\hat{\Lambda}_1)  \right\rangle.\nonumber
\end{eqnarray}

The equation for $\Omega_2$ can be similarly derived as
\begin{eqnarray}
\mathscr{E}_{2^*} \overline{\delta\phi}_{2^*} &=& - \mathscr{A}\mathscr{D}_2 \frac{\omega^*_{0}}{\omega^*_{2}k^2_{r,2}} \partial_r\overline{\delta\phi}_{1}\overline{\delta\phi}_{0^*},\label{eq:DR_2}
\end{eqnarray}
with $\mathscr{E}_2\equiv\mathscr{E}_{EGAM}(\omega_{2^*},k_{r,2^*})$ being the dispersion relation of $\Omega_{2^*}$,  and
\begin{eqnarray}
\mathscr{D}_2\equiv  \left\langle  \frac{\partial_E F_{0,h}}{\omega^*_{0}-\omega_{tr}}  \frac{\omega_{tr}}{\omega^*_{2}-\omega_{tr}} J_0(\hat{\Lambda}_0)J_1(\hat{\Lambda}_0)J_0(\hat{\Lambda}_2) J_1(\hat{\Lambda}_2)  \right\rangle.\nonumber
\end{eqnarray}

 The EGAM two plasmon decay dispersion relation can then be derived  from   equations (\ref{eq:DR_1}) and (\ref{eq:DR_2}) as
\begin{eqnarray}
\mathscr{E}_1\mathscr{E}_{2^*}=&-&\mathscr{A}^2 \frac{|\delta\phi_0|^2 |\omega_0|^2}{\omega_1\omega^*_{2}k_{r,1}k_{r,2}}\mathscr{D}_1\mathscr{D}_2. \label{eq:DR_parametric}
\end{eqnarray}

For the two
stable GBs, we have,
\begin{eqnarray}
\mathscr{E}_1\simeq \mathscr{E}_1 (\omega_1) +\partial_{\omega_1}\mathscr{E}_1 (\omega_1+i\gamma-\omega_0/2)\simeq \frac{4}{\omega_0} (i\gamma+\Delta), \label{eq:DR_1_reduced}
\end{eqnarray}
and similarly,
\begin{eqnarray}
\mathscr{E}_{2^*}\simeq \frac{4}{\omega^*_0}(i\gamma-\Delta),\label{eq:DR_2_reduced}
\end{eqnarray}
with $\Delta\equiv \omega_1-\omega_0/2$ being the   mismatch of half primary mode to GBs, and $\gamma$ being the secondary mode growth rate due to pump  $\Omega_0$ drive. Substituting equations (\ref{eq:DR_1_reduced}) and (\ref{eq:DR_2_reduced}) into   (\ref{eq:DR_parametric}), we have
\begin{eqnarray}
\gamma^2+\Delta^2=-\frac{1}{4} \mathscr{A}^2 |\delta\phi_0|^2\frac{|\omega_0|^2} {k_{r,1}k_{r,2}}\mathscr{D}_1\mathscr{D}_2,\label{eq:DR_parametric_1}
\end{eqnarray}
with the right hand side being the nonlinear EP drive. The secondary modes can be driven unstable as the nonlinear drive overcomes the threshold due to mismatch between half primary mode frequency and GB. Thus, the secondary modes can be excited when the primary amplitude is large enough, or when the frequency mismatch is sufficiently small,  i.e., the secondary mode excitation condition is optimized as the primary mode frequency up sweeping to twice local GAM frequency, as observed in the experiment and numerical simulation \cite{TIdoPRL2016,HWangPRL2018}.

If we consider only resonant EP contribution to the nonlinear coupling,  taking $\omega_{tr,h}\simeq\omega_0$, and assume small EP  drift orbit by taking $J_0(\hat{\Lambda})\simeq 1$, $J_1(\hat{\Lambda})\simeq \hat{\Lambda}/2$,    equation (\ref{eq:DR_parametric_1}) can be reduced to
\begin{eqnarray}
\gamma^2=-\Delta^2+\left(\frac{\pi}{2}\mathscr{A}\right)^2 |\delta\Phi_0|^2 \left\langle \partial_E F_{0,h} \hat{\rho}^2_d \delta (\omega_0-\omega_{tr})\right\rangle^2.\label{eq:DR_parametric_reduced}
\end{eqnarray}
It is clear from equation (\ref{eq:DR_parametric_reduced}) that,   the EPs nonlinearly ``drive" the secondary modes are the same particles that resonantly drive  the primary mode unstable, though the secondary mode frequency is very different from that of the primary mode. Thus, theoretical results from equation (\ref{eq:DR_parametric_reduced}) illuminate experimental observations as well as the findings from numerical simulations. For more quantitative comparison, the threshold of primary EGAM  amplitude  for secondary mode generation can be estimated by
\begin{eqnarray}
|\delta\phi_0|_{threshold}\sim \frac{\mbox{Max($\Delta$, $\gamma_G$)}} {|\pi\mathscr{A}\langle \partial_E F_{0,h} \hat{\rho}^2_d\delta(\omega_0-\omega_{tr})\rangle/2|},
\end{eqnarray}
with $\gamma_G$ being the GAM collisionless damping rate, and $\mbox{Max}(\Delta, \gamma_G)$ giving the maximum value of $\Delta$ and $\gamma_G$.

\section{Conclusions and Discussions}
\label{sec:summary}

In this work, an analytical theory is proposed to interpret the  secondary mode  generation  during primary EGAM frequency chirping  observed  in LHD experiments \cite{TIdoPRL2016}, which is re-produced in MEGA simulation \cite{HWangPRL2018}. The  interpretation is  based on a previous   theory on linear EGAM stability in the same LHD low collisionality plasmas, which shows that for the not fully slowed down EP distribution function, the unstable branch (LBB) frequency   can be significantly higher than the local GAM frequency; while there is a linearly stable branch (GB) with the frequency determined by the local GAM frequency.   The ``primary" and ``secondary" modes in the experimental observations \cite{TIdoNF2015} correspond to the linearly unstable LBB and linearly stable GB, respectively.

It is shown that, the LBB can decay into to two linearly stable GBs as its frequency is up-chirping to twice GB frequency, in a process similar to the well known two plasmon decay process \cite{CSLiuPoF1976,LPowersPoF1984}. The contribution of both thermal plasmas and EPs to the nonlinear process are  derived and evaluated. It is found that  the thermal plasma contribution to the coupling is negligible compared to that of EP FoWs, and this explains   that the nonlinear coupling still occurs   when fluid nonlinearity is turned off in the hybrid MEGA simulation. The resonant EPs play crucial roles in the nonlinear coupling, consistent with  the observation  that  the EPs, which ``drive"  the secondary mode, are the same as those linearly driving the primary mode unstable, though the secondary mode has a frequency much lower than that of the primary.   It is noteworthy that the GB frequency is determined by local GAM frequency, and is only slightly modified when GB and LBB strongly couple. Thus, as the primary frequency keeps chirping up, the secondary mode frequency is almost unchanged, as shown in both LHD experimental observation and MEGA simulation. Thus, the present theory, illuminates  all the crucial evidence from the experiment and simulation, suggesting  this as the mechanism controlling the underlying the physics.

The present theory is   local, facilitated by the existence of the high frequency LBB in the low collisionality condition of  this specific LHD experiment.  However,  two plasmon decay process can also occur in typical discharges  of usual magnetically confined toroidal plasmas, where the unstable EGAM frequency is typically lower than local GAM frequency. This global two plasmon decay process can occur due to the GAM frequency dependence on local plasma parameters, such that EGAM frequency can be significantly higher than GAM continuum frequency of a distant region, if the temperature gradient is   sharp enough. Thus,  the two plasmon decay process can occur as the EGAM tunnels through  the potential barrier, and strongly couple to GAM where the GAM frequency is half of the EGAM. The condition for the above process to happen is, though, quite difficult to satisfy, since 1.  the thermal plasma temperature gradient needs  to be   sharp, such that the potential barrier is narrow enough to have finite EGAM tunneling through, and 2. the characteristic scale length of  thermal plasma temperature  nonuniformity needs to be larger than EP density scale length to have EGAM localized by the potential well. The in-depth discussion of the global coupling process is beyond the scope of the present paper focusing on the specific condition of LHD experiments, and  will be presented in a separated work.

\section*{Acknowledgements}

This work is supported by   the National Key R\&D Program of China  under Grant No. 2017YFE0301900,
the National Science Foundation of China under grant No.  11875233.
This work was also carried out within the framework of the EUROfusion
Consortium and received funding from the EURATOM research and training programme
2014 - 2018 and 2019 - 2020 under Grant Agreement No. 633053 (Project Nos. WP19-ER/ENEA-05 and
WP17-ER/MPG-01). The views and opinions expressed herein do not necessarily reflect
those of the European Commission.

\section*{Data Availability}
The data that support the findings of this study are available from the corresponding author upon reasonable request.

\appendix
\section{Linear stability of EGAM described by equation (\ref{eq:EGAM_DR_linear})}\label{sec:app_A}

Equation (\ref{eq:EGAM_DR_linear}) is the linear EGAM dispersion relation excited by a not fully slowed down EP distribution function given by equation  (\ref{eq:EP_DF}), with the logarithmic and simple-pole like singularities  in the square bracket related to the slowing-down and low energy end cutoff, respectively.  The corresponding EGAM linear properties are controlled by three dominant parameters, i.e., $\omega_b$ and $\omega_L$ determined by NBI birth energy $E_b$ and low energy end cutoff $E_L$, and local GAM frequency $\omega_G$.

\begin{figure}[h]
\includegraphics[width=8cm]{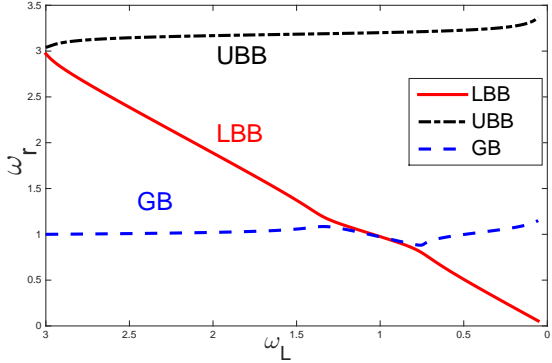}
\caption{Real frequencies of the three modes dependence on $\omega_L$ with  $\omega_b=3\omega_G$. The frequencies are in unites of $\omega_G$. }\label{fig:RF_SPA}
\end{figure}

\begin{figure}[h]
\includegraphics[width=8cm]{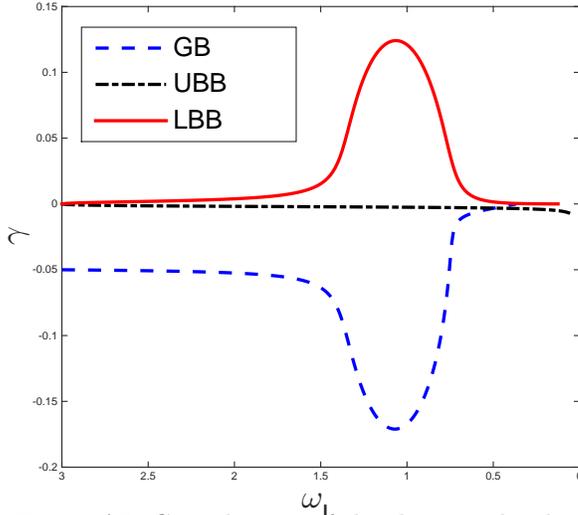}
\caption{Growth rates of the three modes dependence on $\omega_L$ with  $\omega_b=3\omega_G$.}\label{fig:GR_SPA}
\end{figure}

Here, equation (\ref{eq:EGAM_DR_linear}) was solved for the EGAM stability v.s. $\omega_L$,  given $\omega_b=3\omega_G$ and $\lambda_0B_0<2/5$, such that only the simple pole like singularity is destabilizing. Besides, a small but finite GAM damping rate is assumed $\gamma_G=-0.05\omega_G$.  Note that  $\omega_L=\omega_b$ corresponds to all the EPs having the same energy; i.e.,  to beam ions having not slowed down at all; while $\omega_L\ll\omega_b$ (more precisely, $E_b>E_L\gg E_{crit}$) corresponds to NBI being fully slowed down.  The dependence of the  real frequencies of the three branches on $\omega_L$ is given in Fig. \ref{fig:RF_SPA}, while their growth rates are given in Fig. \ref{fig:GR_SPA}, with the frequencies/growth rates normalized with $\omega_G$.  It is shown that, UBB frequency is determined by $\omega_b$  while it remains almost independent of  $\omega_G$ or $\omega_L$,  and it is marginally stable.   The   LBB frequency is determined by $\omega_L$,  and it can be unstable even when its frequency is significantly higher than local GAM frequency.  The GB frequency is determined by local GAM frequency, and is almost independent of $\omega_L$.

The important information here is that, 1. LBB frequency is determined by $\omega_L\equiv\sqrt{2E_L(1-\lambda_0B_0)}/(qR_0)$ and the unstable LBB frequency can be significantly higher than local GAM frequency if strongly driven; and 2. GB frequency is determined by local GAM frequency, and its dependence on $\omega_L$ is very weak.  These two points are crucial for the interpretation of the nonlinear process in Sec. \ref{sec:two_plasmon_decay}.

\section{Self-consistent EGAM frequency chirping due to pitch angle scattering}
\label{sec:app_B}

The nonlinear evolution of EGAM including   frequency chirping, can be derived by self-consistently including the slowly evolution of the ``equilibrium" EP distribution function $F_{0,h}$ due to the nonlinear interactions with EGAM \cite{FZoncaNJP2015,FZoncaJPCS2021}, as addressed in Ref. \cite{ZQiuPST2018}. The corresponding $F_{0,h}$ evolution  obeys the following Dyson equation \cite{FZoncaNJP2015,FZoncaJPCS2021,MFalessiNJP2021,CItzyksonbook1993}
\begin{eqnarray}
\hspace*{-2em}\hat{\omega} \hat{F}_{0,h} &=& -\frac{e^2\hat{\omega}_d}{16} |\delta\phi_G|^2 \frac{\partial}{\partial E}\left[ \frac{\hat{\omega}_d (\hat{\omega}-i\gamma)}{(\hat{\omega}-i\gamma)^2-(\omega_{0,R}-\omega_{tr,h})^2}\right] \frac{\partial}{\partial E} \hat{F}_{0,h}(\hat{\omega}-2i\gamma)\nonumber\\
&& + i F_{0,h}(0). \label{eq:EGAM_dyson}
\end{eqnarray}
Here, $\hat{\omega}$ denotes the slow nonlinear evolution of $F_{0,h}$ from its initial value $F_{0,h}(0)$, $\hat{F}_{0,h}$ is the Laplace transform of $F_{0,h}$, and $\omega_{0,R}$ is the real frequency of EGAM.  Equation (\ref{eq:EGAM_dyson}) describes the self-consistent evolution of $F_{0,h}$, due to emission and re-absorption of a single coherent EGAM. In deriving equation (\ref{eq:EGAM_dyson}), only evolution in $E$ (or $v_{\parallel}$ or $\lambda$) is considered, since both $P_{\phi}$ and $\mu$ are well conserved for GAM/EGAM with $n=0$ and frequency significantly lower than $\Omega_i$. Equation (\ref{eq:EGAM_dyson}) is derived assuming well circulating EPs and $\hat{\Lambda}_k\ll1$, while a more systematic treatment can be done following Ref. \cite{FZoncaNJP2015}.

EGAM may scatter EPs to smaller pitch angle, and thus, larger $\omega_{tr,h}$ regime, which will lead to self-consistent EGAM frequency up-chirping as its frequency is nonperturbatively determined by $\omega_L$, as we shown in  \ref{sec:app_A}. However, the self-consistent EGAM evolution will be only qualitatively but not  quantitatively investigated  in the present work, since it will be addressed in an independent work.

\section*{References}

\providecommand{\newblock}{}

\end{document}